\RequirePackage{etoolbox}
\csdef{input@path}{{style/}{graphics/}}
\documentclass[aoas,MSNbibl,nameyear,dvips]{arximspdf}
\makeatletter
   \@ifpackageloaded{graphicx}{}{\usepackage{graphicx}}
\makeatother
\usepackage{dcolumn}
\usepackage{graphicx}

\doi{10.1214/14-AOAS795}
\volume{9}
\issue{1}
\pubyear{2015}
\firstpage{200}
\lastpage{224}
\docsubty{FLA}

\makeatletter
\newcolumntype{d}[1]{D{.}{.}{#1}}

\makeatother

\begin{document}
\begin{frontmatter}

\title{Estimating the relative rate of recombination to mutation in
bacteria from single-locus variants using composite likelihood
methods\thanksref{T1}}
\runtitle{Recombination in bacteria}
\thankstext{T1}{Supported in part by the Marsden Fund Project 08-MAU-099 (Cows, starlings and Campylobacter in New Zealand:
unifying phylogeny, genealogy and epidemiology to gain insight into pathogen evolution).}

\begin{aug}
\author[A]{\fnms{Paul}~\snm{Fearnhead}\corref{}\thanksref{T2,m1}\ead[label=e1]{p.fearnhead@lancaster.ac.uk}},
\author[B]{\fnms{Shoukai}~\snm{Yu}\thanksref{m2}\ead[label=e2]{S.Yu1@massey.ac.nz}},
\author[B]{\fnms{Patrick}~\snm{Biggs}\thanksref{m2}\ead[label=e3]{P.Biggs@massey.ac.nz}},
\author[C]{\fnms{Barbara}~\snm{Holland}\thanksref{m3}\ead[label=e4]{Barbara.Holland@utas.edu.au}}
\and
\author[B]{\fnms{Nigel}~\snm{French}\thanksref{m2}\ead[label=e5]{N.P.French@massey.ac.nz}}
\thankstext{T2}{Supported in part by the Engineering and Physical Sciences Research Council, UK, Grant EP/K014463/1.}
\runauthor{P. Fearnhead et al.}
\affiliation{Lancaster University\thanksmark{m1}, Massey
University\thanksmark{m2} and University of Tasmania\thanksmark{m3}}
\address[A]{P. Fearnhead\\
Department of Mathematics and Statistics\\
Fylde College\\
Lancaster University\\
Lancaster LA1 4YF\\
United Kingdom\\
\printead{e1}}
\address[B]{S. Yu\\
P. Biggs\\
N. French\\
Infectious Disease Research Centre\\
Institute of Veterinary, Animal\\
\quad and Biomedical Sciences\\
Private Bag 11 222\\
Massey University\\
Palmerston North 4442\\
New Zealand \\
\printead{e2}\\
\phantom{E-mail:\ }\printead*{e3}\\
\phantom{E-mail:\ }\printead*{e5}}
\address[C]{B. Holland\\
School of Mathematics and Physics\\
University of Tasmania\\
Hobart\\
Tasmania 7001\\
Australia\\
\printead{e4}}
\end{aug}

%
\received{\smonth{9} \syear{2013}}
%
\revised{\smonth{6} \syear{2014}}

%
\begin{abstract}
A number of studies have suggested using comparisons between DNA
sequences of closely related bacterial isolates to estimate the relative
rate of recombination to mutation for that bacterial species. We
consider such an approach which uses single-locus variants: pairs of isolates
whose DNA differ at a single gene locus. One way of deriving point
estimates for the relative rate of recombination to mutation from such
data is to use composite likelihood methods. We extend recent work in
this area so as to be able to construct confidence intervals for our estimates,
without needing to resort to computationally-intensive bootstrap
procedures, and to develop a test for whether the relative rate varies
across loci.
Both our test and method for constructing confidence intervals are
obtained by modeling the dependence structure in the data,
and then applying asymptotic theory regarding the distribution of
estimators obtained using a composite likelihood.
We applied these methods to multi-locus sequence typing (MLST) data
from eight bacteria, finding strong evidence for considerable rate
variation in three of these: \emph{Bacillus cereus}, \emph{Enterococcus
faecium} and \emph{Klebsiella pneumoniae}.
\end{abstract}

%
\end{frontmatter}

\section{Introduction}\label{sec1}

Homologous recombination is a process which allows foreign DNA to be
incorporated within a genome. In bacteria this can occur through three
different mechanisms: conjugation, the uptake of DNA from other
bacteria; transformation, the uptake of naked DNA from the remains of
bacteria that exist in the living environment; or transduction, where
DNA is implanted by bacteriophage [\citet{Low:1978}]. Although
different, each
result in the \hyperref[sec1]{Introduction} of a new DNA sequence within a region of the
genome, and thus recombination is potentially an important mechanism
driving the evolution of a given bacteria. Understanding recombination
in bacteria
is important because it can allow for genetic exchange between distant
bacterial species and impacts on the evolution of new species [\citet
{Fraser:2007,Sheppard:2008}].
Furthermore, the rate of recombination varies considerably across
bacterial species: with estimates of the relative effect of
recombination to mutation varying by over three orders of magnitude in
\citet{Vos/Didelot:2009}. Here we look at how
to estimate the relative rate of recombination to point mutation from
population genetic data that describe the genetic variation between a
sample of bacterial isolates at a number of loci. In particular, our
approach develops recent ideas
that estimate this relative rate by comparing the DNA for closely
related isolates.

For population genetic data it is often helpful to consider the
genealogical history of a sample. If there is no recombination, this
can be represented by a single tree, which is often called the
genealogy of the
sample. The effect of recombination is that, while at any specific
position along the chromosome we can define such a genealogical tree,
this tree can be different for different positions. The genealogical
history of
a sample is thus defined by the collection of all such trees, which can
be represented through a graph [\citet{Griffiths/Marjoram:1996b}].

\begin{figure}

\includegraphics{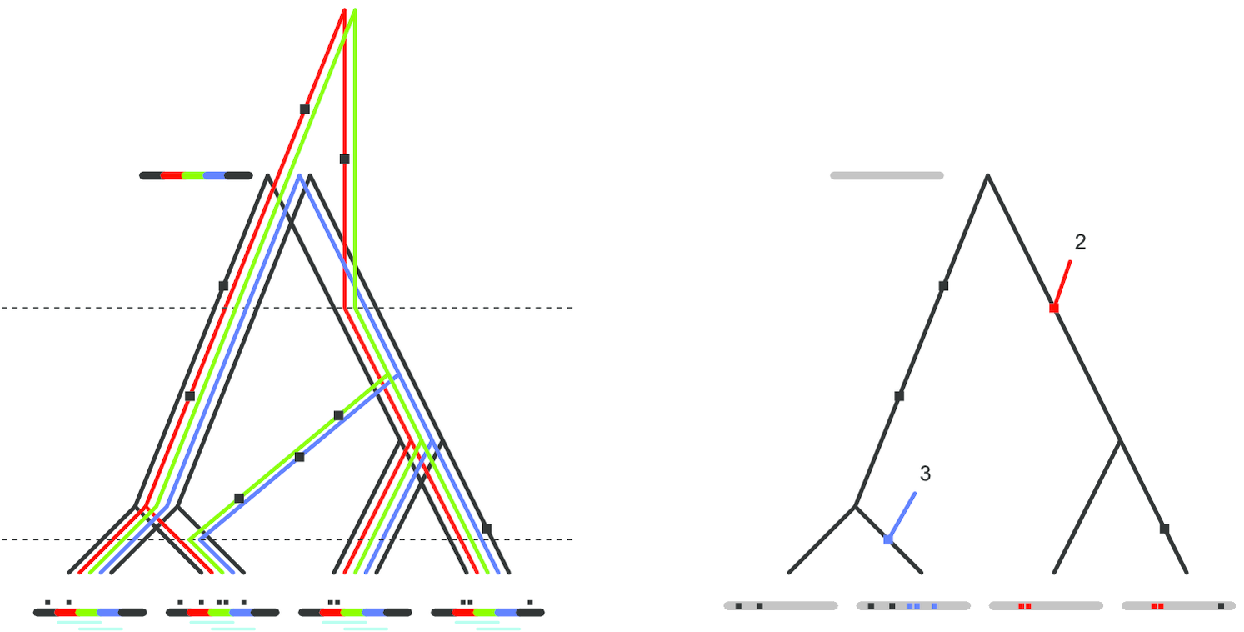}

\caption{Left-hand plot: Example of the genealogical
trees for 4 isolates in one region of the genome.
The effect of recombination is that different sites in the gene have
different genealogical trees.
Here we have two recombination events that have affected the
genealogical relationship of the isolates
(the position of these back in time are denoted by the dashed lines,
and the regions they affect by the light blue lines under the gene fragments).
The black tree is the clonal frame of the sample: the genealogy at
regions unaffected by recombination. Mutations that have affected the
sample are given by the black squares. Here we consider mutations that
create differences from the sequence of the common ancestor on the
clonal frame. The other trees represent the genealogies of regions
affected by either one or both recombination events. Right-hand plot:
the simplified model for the data based on the clonal frame.
Differences within our sample are created by mutation and recombination
events that occur on the clonal frame.
We do not track the ancestry of recombination events, instead each one
just introduces a number of mutations within the recombinant region.
These events are shown
in the figure and are labeled with the number of differences introduced.}\label{Fig:1}
\end{figure}

Within bacteria each recombination event generally affects only a
relatively small region of the genome. For example, in \textit{Campylobacter jejuni}, a recombination event
may change the DNA within a region of between a few hundred to a few
thousand base pairs, which constitutes a fraction of a percent of the
whole 1.6~Mb genome. Thus, we can define a single tree for a sample
of bacteria by tracing the ancestry of the nonrecombinant region at
each recombination event. This tree has been called the clonal frame
[\citet{Milkman:1990,Didelot/Falush:2007}]. We can then model recombination
events as introducing a number of mutations onto this clonal frame. An
example is given in Figure~\ref{Fig:1}. Our approach to estimating
recombination rates in this paper is based on using such a model.

In this paper we assume we have genetic data collected from a number of
isolates of a given bacteria, and that this genetic data
consist of the DNA sequence at $L$ loci of similar size. We assume
these loci are sufficiently spread around the genome such that
a single recombination event is unlikely to affect more than one locus.
An example of data satisfying these
assumptions is MLST data [\citet{Maiden:1998}], which consist of the DNA
sequence of $\approx$500~bp fragments from a selection of, normally around
7, housekeeping genes. Large MLST data sets for over 20 bacteria are
available from {\url{http://pubmlst.org}}.

\begin{figure}

\includegraphics{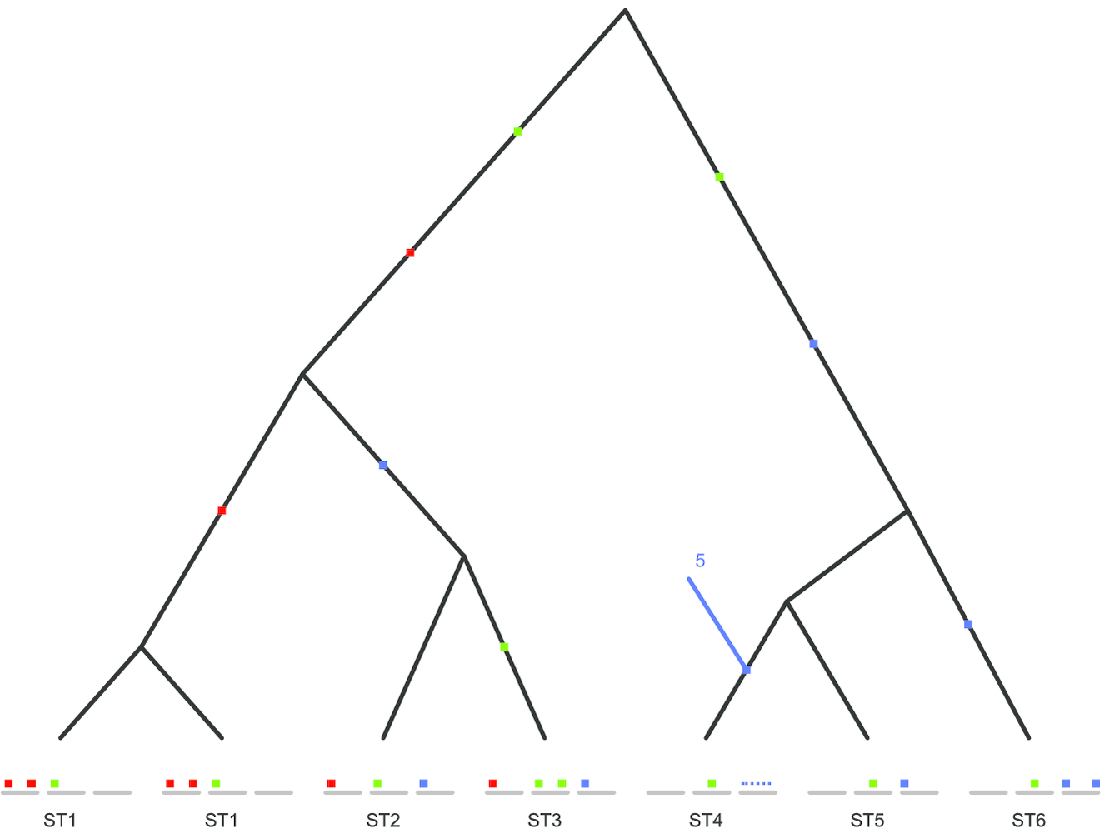}

\caption{Example clonal frame and resulting data set for
7 isolates at 3 loci. Mutations are denoted by squares, and the color
denotes the locus that the mutation occurs on. There is a single
recombination event, denoted by
a square labeled with the number of differences to the ancestral
sequence that event introduces. For this example, there are 6
distinct sequence types, denoted ST1 to ST6. We have an SLV pair at
locus 2 (ST2 and ST3) and 3 SLV pairs at locus 3 (all distinct pairs
of ST4 to ST6).}\label{Fig:2}
\end{figure}

For such data we can define \emph{sequence types (STs)} so
that two isolates which have identical sequences at all $L$ loci will
have the same ST, but any two isolates whose sequences differ
will have differents STs. It is standard to define STs numerically:
ST1, ST2 etc. A simple example for 3 loci and 7 bacterial isolates is
given in Figure~\ref{Fig:2}, where we also show the underlying clonal
frame of the
sample, and the
mutation and recombination events that have affected the sample. If we
assume each mutation is distinct, and these are also different from the
mutations introduced at recombination events, then we get 6 distinct
sequence types.

The single-locus variants (SLVs) of a specific ST will be the set of
other STs that differ from it only at a single-locus. If we consider
pairs of SLVs at a specific locus $l$, then an SLV pair will be defined
as a pair of isolates that have different DNA sequences at locus $l$
but have
identical DNA sequences at the other $L-1$ loci. For the example in
Figure~\ref{Fig:2} we get one SLV pair at locus 2, and 3 SLV pairs at
locus 3. These can be summarized by
the STs of each pair together with the number of nucleotide differences
at the locus that differs; see Table~\ref{Tab:data}. In this paper we
consider how to
use data such as that in Table~\ref{Tab:data} to infer the relative
rate of recombination to mutation. Note that we define this relative
rate as the ratio of the rate at which a locus is affected
by recombination to the rate at which it is affected by mutation. This
rate is different from the relative rate of recombination to mutation
events across the genome, as recombination events that
start outside a locus can still affect it. Thus, the rate at which a
locus is
affected by recombination depends both on the rate of recombination
events and the relative size of the average recombination tract length
to the size of the locus. 

\begin{table}[b]
\tablewidth=200pt
\caption{Data for SLV pairs from example in Figure
\protect\ref{Fig:2}. The data consist of the SLV pairs for each locus, together
with $x$, the number of nucleotide differences each SLV pair has at
that locus}\label{Tab:data}

\begin{tabular*}{200pt}{@{\extracolsep{\fill}}lccc@{}}
\hline
\textbf{Locus} & \textbf{ST} & \textbf{ST} &$\bolds{x}$ \\
\hline
2 & 2 & 3 & 1 \\
3 & 4 & 5 & 5 \\
3 & 4 & 6 & 6 \\
3& 5 & 6 & 1 \\
\hline
\end{tabular*}
%
\end{table}

The idea of using SLVs to estimate the relative rate of recombination
to mutation comes initially from the work of \citet{Feil:1999} [see also
\citet{Feil:2000,Spratt:2001}]. By comparing closely related isolates,
it can often be clear as to whether the pair of isolates
differs only by mutation or not. The approach in \citet{Feil:1999} is
based on assuming that if the number of nucleotide differences is small
(say 2 or fewer), then these are caused by mutation. If the number of
differences is large, then these
are caused by recombination. For the data in Table~\ref{Tab:data} such
an approach would work well: identifying correctly that two of the four
SLV pairs are created by mutation, and two involve recombination.

However, to obtain sensible estimates of the relative rate of
recombination to mutation, we need to deal with two issues. First is
the fact that SLV pairs that involve recombination may have differences
caused by both mutation and
recombination. Thus, using a simple ratio of SLV pairs caused by only
mutation to those that involve recombination may not be appropriate.
Second, some recombination events may introduce a small number of
nucleotide differences,
and thus we need to allow for some of the SLV pairs that differ at a
small number of nucleotides to be due to recombination. To address
these issues, \citet{Yu:2012} introduce a simple model for the number of
nuceotide differences for an SLV pair as
a function of the relative rate of recombination to mutation, and then
estimate the parameter in this model using composite likelihood. We
take the same approach here.

While the model is approximate, it should give robust and accurate
inferences in situations where it is easy to detect whether SLVs are
caused only by mutation and where it is likely that SLVs are caused by
only one, mutation
or recombination, event. This corresponds to SLVs defined for data
collected at a relatively large number of loci (such as the 7 used in
MLST data), and where most recombination events introduce a large
number of nucleotide
differences.

A second issue with using SLVs to estimate the relative rate of
recombination to mutation is that data from some SLV pairs are
dependent. This can be seen in the data in Table~\ref{Tab:data} and the
three SLV pairs at locus 3.
These three SLV pairs are caused by two events: one mutation and one
recombination. This dependence makes it harder to assess uncertainty in
parameter estimates. As a result, \citet{Yu:2012} resort to using
simulation, via a
parametric bootstrap, to assess this uncertainty. The main disadvantage
with this is that the accuracy of the resulting measures of uncertainty
will depend on how accurate the model used to simulate the data is and,
in practice,
the models used to simulate data do not capture many of the features
observed in real data sets. Furthermore, the use of simulation adds
substantially to the overall computational effort required to analyze
any data set: using a
parametric bootstrap for large data sets, such as the \emph
{Staphylococcus aureus} MLST data set we analyze in Section~\ref{S:Bac},
can take months of CPU time.

The main difference between this work and that of \citet{Yu:2012} is
that we use theory for composite likelihoods to get direct measures of
uncertainty of our estimates. We believe these to be more reliable than using
simulation, and they are much quicker to calculate.
Composite likelihood theory also gives a framework for performing
inference across loci. We show how we can test whether there are
differences in the relative rate
of recombination to mutation across loci, and how to get an estimate of
the common relative rate under the assumption that there is no
variation across loci.

The next section introduces our model for data from SLV pairs and how
we can use composite likelihood to estimate the relative rate of
recombination to mutation from such data and to
quantify the uncertainty in these estimates. We evaluate this method on
simulated data in Section~\ref{S:SS}. Our results suggest that we can get both accurate estimates
and also appropriate measures of uncertainty of our estimators. For
large data sets, coverage values of our confidence intervals do drop
away from their nominal value, but this appears
to be due to slight biases in our model as opposed to underestimating
the variability of the estimators. We show that tests for detecting
differences across loci have close to their nominal significance level
across a range
of simulated scenarios, and have good power to detect rate variation
across loci in situations where the recombination varies by a factor of
three or more. In Section~\ref{S:Bac} we apply our method to analyze
MLST data from 8 bacteria. In three cases we find strong evidence
that the rate of recombination to mutation varies across loci, with
estimates suggesting this variation could be by up to two orders of
magnitude. The paper ends with a discussion. Data and code used in the
paper are available at
{\url{http://www.maths.lancs.ac.uk/\textasciitilde fearnhea/SLV.zip}}.

\section{Estimating recombination rates from SLVs}

We will now describe how we use data like that in Table~\ref{Tab:data}
to estimate the relative rate of recombination and mutation at a locus
using composite likelihood methods. Roughly, the idea behind composite
likelihood
approaches is to (i) split data into small subsets; (ii) introduce a
probabilistic model for each subset of data, which in turn will define
a log-likelihood for that subset; and (iii) combine information from
all subsets through
taking a weighted sum of the log-likelihoods from the subsets. The
weighted sum is called a composite likelihood. Parameters can be
estimated through maximizing this composite likelihood. Furthermore, if we
model the dependence between the log-likelihoods across different
subsets, we are able to estimate the asymptotic variance of, and
construct confidence intervals for, the resulting parameter estimates.

Implementing a composite likelihood method involves a number of key
decisions and modeling assumptions. The first is the choice of subsets
of the data to use. In our application each subset corresponds to an
SLV pair, with the
data being the number of nucleotide differences we observe for that SLV
pair. Second, we need to develop a model for the data, and we describe
our model in Section~\ref{S:SLVpair}. Then we need to choose the weights
used when constructing the composite likelihood, and finally to choose
how to model the dependence between subsets of the data. The last
aspect is needed in order to be able to assess the uncertainty of estimators,
and hence to obtain confidence intervals. One advantage of composite
likelihood methods is that we need only model one
aspect of this dependence, namely, the covariance of the score function
(the derivative of the log-likelihood) for pairs of subsets of data.
Our choices of weights and model for this dependence for our
application are
introduced in Section~\ref{S:CL}, together with fuller details of how
we then construct confidence intervals for our estimate of the relative
rate of recombination and mutation.

Theory for composite likelihood can also be used to
combine information across loci and construct a test for whether the
relative rate of recombination to mutation varies across loci. This is
described in Section~\ref{S:Joint}.

\subsection{Model for a single SLV pair}\label{S:SLVpair}

Here we derive a conditional likelihood for data from a single SLV pair
at a specific locus, $l$, say. The data are the number of nucleotide
differences that the SLV pair has at that locus, which will be denoted~$x$.
Remember that the total number of loci is $L$, and we will denote the
relative rate at which recombination, as opposed to mutation, affects
locus $l$ by $\lambda$. Our conditional log-likelihood will be based on
the log of the
probability of observing $x$ differences between two STs conditonal on
these STs being an SLV pair at locus $l$,
\[
\ell(\lambda;x)=\log\Pr_{\lambda}(X=x|\mbox{SLV at locus $l$}),
\]
where the random variable $X$ denotes the number of nucleotide
differences at locus $l$ between a (random) pair of isolates, and we
are conditioning on the pair of isolates being an SLV pair at locus
$l$. We include
the subscript $\lambda$ as the probability depends on $\lambda$. 

To calculate this conditional probability, let $A$ denote the event
that only mutations have occurred at locus $l$ to create the
differences between the pair of isolates, and $A^c$ the complementary
event that at least one
recombination occurred at locus~$l$. Further, let $\theta_l$ denote the
mutation rate at locus $l$,
and $\theta=\sum_{i=1}^L \theta_i$ denote the overall mutation rate
across the $L$ loci. [We use standard coalescent scaling for these
rates, so
one unit of time is equal to the expected time in the past until a pair
of isolates share a common ancestor; see \citet{Wakeley:2007}.]
Finally, we shall assume that the relative rate of recombination to
mutation, $\lambda$, is the same across all loci.

Under a standard coalescent model, the probability of a pair of
isolates being an SLV pair at locus $l$ is
\begin{eqnarray*}
\Pr_\lambda(\mbox{SLV at locus $l$})&\approx&\frac{1}{1+(1+\lambda
)\theta} \sum
_{i=1}^\infty \biggl(\frac{(1+\lambda)\theta
_l}{1+(1+\lambda
)\theta}
\biggr)^i
\\
&\approx& \frac{1}{(1+\lambda)\theta} \sum_{i=1}^\infty
\biggl(\frac
{(1+\lambda)\theta_l}{(1+\lambda)\theta} \biggr)^i
\\
&=& \frac{1}{\theta(1+\lambda)}\frac{\theta_l}{\theta} \biggl(1-\frac
{\theta_l}{\theta}
\biggr)^{-1}= \frac{1}{\theta(1+\lambda)}\frac
{\theta
_l}{\theta-\theta_l}.
\end{eqnarray*}
%
The first approximation comes from assuming that all recombination
events at a locus introduce a change to the sequence of one of the
isolates. The expression on the right-hand side is then
obtained by considering the possible events in the history of the two
isolates back to their common ancestor: there will need to be at least
one mutation/recombination event at locus $l$,
and no mutation/recombination events at other loci. The
$1/(1+(1+\lambda
)\theta)$ is the probability that the next event back in time is a
coalescent event, and the $(1+\lambda)\theta_l/(1+(1+\lambda)\theta)$
is the
probability of the next event back in time being a mutation or
recombination event at locus $l$ [\citet{Wakeley:2007}].
We sum over $i$, the number of mutation or recombination
events in the history of the SLV pair. The approximation we have used
in the second line is reasonable if $(1+\lambda)\theta\gg 1$, which will
be true for the cases where this approach
to inference for $\lambda$ can be expected to be accurate. The
advantage of this approximation is that it means our final expression
for the conditional likelihood (see below)
will only depend on $\theta_l$ and $\theta$ through
the ratio $\theta_l/\theta$, which will be easier to estimate in practice.

Now, by a similar argument, and using the same approximation, we can get
\[
\Pr_\lambda(X=x \cap A \cap\mbox{SLV at locus $l$}) \approx
\frac
{1}{\theta(1+\lambda)} \biggl( \frac{\theta_l}{\theta(1+\lambda)} \biggr)^x,
\]
as this will require $x$ mutation events at locus $l$ followed by a
coalescent event. Summing over $x=1,2,\ldots$ gives
\[
\Pr_\lambda(A\cap\mbox{SLV at locus $l$})\approx\frac{1}{\theta
(1+\lambda)}
\frac{\theta_l}{\theta+\lambda\theta-\theta_l}.
\]
Thus, we can write
\begin{eqnarray*}
&&\Pr_{\lambda}(X=x|\mbox{SLV at locus $l$}) \\
&&\qquad\propto\Pr
_\lambda (X=x \cap\mbox{SLV at locus $l$})
\\
&&\qquad= \Pr_\lambda(X=x \cap A \cap\mbox{SLV at locus $l$}) + \Pr
_\lambda \bigl(X=x \cap A^c \cap\mbox{SLV at locus $l$}\bigr)
\\
&&\qquad\approx \frac{1}{\theta(1+\lambda)} \biggl( \frac{\theta
_l}{\theta
(1+\lambda)} \biggr)^x \\
&&\qquad\quad{}+
\Pr_\lambda\bigl(\mbox{SLV at locus $l$} \cap A^c\bigr)
\Pr_\lambda\bigl(X=x|\mbox{SLV at locus $l$},A^c\bigr)
\\
&&\qquad= \frac{1}{\theta(1+\lambda)} \biggl( \frac{\theta_l}{\theta
(1+\lambda
)} \biggr)^x\\
&&\qquad\quad{} + \Bigl(
\Pr_\lambda(\mbox{SLV at locus $l$}) - \Pr _\lambda (\mbox{SLV at
locus $l$} \cap A) \Bigr)\\
&&\qquad\quad{}\times\Pr_\lambda\bigl(X=x|\mbox{SLV at locus
$l$},A^c\bigr)
\\
&&\qquad\approx \frac{1}{\theta(1+\lambda)} \biggl( \frac{\theta
_l}{\theta
(1+\lambda)} \biggr)^x
\\
& &\qquad\quad{}+ \biggl(\frac{1}{\theta(1+\lambda)}\frac{\theta_l}{\theta
-\theta_l} - \frac{1}{\theta(1+\lambda)}
\frac{\theta_l}{\theta+\lambda
\theta
-\theta_l} \biggr)\\
&&\qquad\quad{}\times\Pr_\lambda\bigl(X=x|\mbox{SLV at locus
$l$},A^c\bigr)
\\
&&\qquad\propto \biggl( \frac{\theta_l}{\theta(1+\lambda)} \biggr)^x+ \biggl(
\frac{\theta_l}{\theta-\theta_l} - \frac{\theta_l}{\theta
+\lambda\theta
-\theta_l} \biggr)\Pr_\lambda\bigl(X=x|
\mbox{SLV at locus $l$},A^c\bigr).
\end{eqnarray*}
We can calculate the normalizing constant of this distribution by
summing the final terms over all possible values for $x$.

To use this conditional likelihood, we need to specify the
probabilities $\Pr_\lambda(X=x|\mbox{SLV},A^c)$ and $\theta
_l/\theta$.
We approximate the former by the distribution of the number of
nucleotide differences
that are introduced by a single recombination event, on the basis that
we expect the number of mutation/recombination events in the history of
an SLV pair to be small, and the recombination
event will produce most of the differences between the two isolates.
This can then be empirically
approximated based on simulating recombination events at locus $l$ from
data on the population diversity of sequences at that locus (see
Appendix \ref{appa}). To estimate $\theta_l/\theta$, we can either use the
relative size of the sequences at
each locus (approximating the mutation rate per bp as constant) or we
can use estimates of the mutation rate at each locus from population
data, such as based on the mean number of pairwise differences [\citet
{Donnelly/Tavare:1995}]. Our experience
is that the results are robust to the method used, and we suggest the
former unless there is strong evidence that mutation rates vary
substantially across loci.

This conditional likelihood is very similar to that derived in \citet
{Yu:2012}. The main difference is that a further approximation is used
in \citet{Yu:2012} whereby the probability of two isolates being an SLV
pair does
not depend on $\lambda$.

\subsection{Composite likelihood inference} \label{S:CL}

We now consider how to estimate $\lambda$ given data from a set of $n$
SLV pairs. Denote the number of differences at each of the $n$ SLV
pairs by $\mathbf{x}=(x_1,\ldots,x_n)$.
If the data from each SLV pair were independent from the others, then
it would be natural to estimate $\lambda$ by maximizing
$\sum_{i=1}^n \ell(\lambda;x_i)$;
however, the data from each SLV pair is not necessarily independent. To
see this, consider the SLV pairs in Table~\ref{Tab:data}. We have three
SLV pairs involving each pair of ST4, ST5 and ST6.
The data for these three SLV pairs are dependent, as different SLV
pairs are affected by the same events. For example, both SLV pairs
involving ST4 are affected by the same recombination event.

Despite the data being dependent, we can still estimate $\lambda$ by
maximizing a related function
\[
\operatorname{Cl}(\lambda;\mathbf{x})=\sum_{i=1}^n
w_i \ell(\lambda;x_i),
\]
where $w_1,\ldots,w_n$ are a set of positive weights. In this case
$\operatorname{Cl}(\lambda;\mathbf{x})$ is often called a composite likelihood,
and asymptotic theory exists which shows that in many situations
maximizing a composite likelihood produces an estimator with good
statistical properties, such as consistency and asymptotic normality
[\citet{Varin:2011}].
The choice of weights affects the overall accuracy of the estimator,
and we suggest an appropriate choice for our application below.
Furthermore, this theory shows how to construct appropriate confidence
intervals for
parameters. Here we outline one such result that we will use.

Assume the true parameter value is $\lambda_0$ and that $\hat\lambda$
maximizes $\operatorname{Cl}(\lambda;\mathbf{x})$. Define the score function
\begin{eqnarray*}
u(\lambda;x)&=&\frac{\mathrm{d} \ell(\lambda;{x})}{\mathrm{d}\lambda}\quad\mbox{and}
\\
U(\lambda;\mathbf{x})&=&\frac{\mathrm{d} \operatorname{Cl}(\lambda;\mathbf
{x})}{\mathrm
{d}\lambda}=\sum_{i=1}^n
w_i u(\lambda;x_i).
\end{eqnarray*}
Define $J(\lambda)=\operatorname{Var}(U(\lambda;\mathbf{X}))$ and
\[
I(\lambda)=-\mathrm{E} \biggl(\frac{\mathrm{d}U(\lambda;\mathbf
{X})}{\mathrm
{d}\lambda} \biggr),
\]
where in each case we calculate the variance or expectation with
respect to data sets $\mathbf{X}$ being drawn from the model with
parameter $\lambda$.
Then if we set $\gamma=J(\lambda_0)/I(\lambda_0)$, we can calculate a
scaled deviance
\[
W(\lambda)=\frac{2}{\gamma} \bigl[\operatorname{Cl}(\hat\lambda;\mathbf {x})-\operatorname{Cl}(\lambda ;
\mathbf{x}) \bigr].
\]
Under certain regularity conditions, as $n\rightarrow\infty$,
$W(\lambda
_0)$ is asymptotically chi-squared distributed with 1 degree of
freedom. If we can calculate, or consistently estimate $\gamma$, then this
result can be used to construct a confidence interval for $\lambda$.
Note that if $\operatorname{Cl}(\lambda;\mathbf{X})$ is replaced by a true
log-likelihood, then, under standard regularity conditions, both
$I(\lambda)$ and $J(\lambda)$
are equal to the Fisher information. In this case $\gamma=1$.

Each $u(\lambda;X_i)$ is identically distributed. Let $\sigma^2=\operatorname
{Var}(u(\lambda_0;X))$, and note that standard results for the expected
information give that $I(\lambda_0)=\break \sum_{i=1}^n w_i \sigma^2$. Now we
can write
%
\begin{equation}
\label{eq:J} J(\lambda_0)=\sigma^2 \Biggl(\sum
_{i=1}^n \sum_{j=1}^n
w_iw_j\operatorname {Cor}\bigl[u(\lambda;X_i);u(
\lambda;X_j)\bigr] \Biggr).
\end{equation}

For any data set we will be able to partition the STs involved in SLV
pairs at locus $l$ into groups, so that if you take any pair of STs
within a group, they will form an SLV pair, but if you take any two STs
which come from
different groups, they will not form an SLV pair. Assume that there are
$G$ groups, containing $n_1,\ldots,n_G$ STs, respectively.

A consequence of this is that you can also split the set of SLV pairs
into the same number of groups: so the $g$th group of SLV pairs will
consist of all $n_g(n_g-1)/2$ pairs from the $g$th group of STs. The
total number of SLV
pairs will be $n=\sum_{g=1}^G n_g(n_g-1)/2$.
The dependence in the data, the number of nucleotide differences of
each SLV pair, is due to the possibility of different SLV pairs being
affected by the same mutation and/or recombination events, as was seen
in the example
in Table~\ref{Tab:data}. By construction, SLV pairs taken from
different groups will not share any mutation or recombination events,
hence, it is reasonable to assume the data from SLV pairs in different groups
will be independent. For all distinct SLV pairs within the same group,
the correlation between them will be the same (by symmetry). To develop
a parsimonious model for the correlation structure within a group, we
will make
a number of further simplifying assumptions. First is that the
correlation is the same for distinct SLV pairs as it is for SLV pairs
that share a common ST. Second is that the level of correlation does
not depend on
the size of the group. Under these assumptions we will have for some
$\alpha\in[0,1]$,
\begin{eqnarray*}
&&\operatorname{Cor}\bigl[u(\lambda;X_i);u(\lambda;X_j)
\bigr]\\
&&\qquad=\cases{ %
1, &\quad  $\mbox{if $i=j$},$
\vspace*{2pt}\cr
\alpha,&\quad $\mbox{if $i\neq j$ but SLV pairs}
\mbox{ $i$ and $j$ are in the same group},$
\vspace*{2pt}\cr
0, & \quad $\mbox{otherwise.}$}
\end{eqnarray*}

We choose the weights based on this dependence structure. A group of
SLV pairs based on $n_g$ STs will contribute $n_g(n_g-1)/2$ terms to
the composite likelihood, but these depend on just $n_g$ pieces of
information (the data
at the $n_g$ different STs). If we chose uniform weights, then this
would mean that the composite likelihood could be overly dominated by
the data from a small number of large groups, which would lead
to an increase of the variance of our estimate of $\lambda$. Thus, we
choose to downweight SLV pairs that are in large groups. Our particular
choice is that for an SLV pair in a group of size $n_g$
we have $w_i=f_w(n_g)=\{n_g(n_g-1)/2\}^{-1/2}$, the inverse of the
square root of the number of SLV pairs in that group.

Substituting these definitions for the correlation and the weights into
(\ref{eq:J}) gives
\begin{eqnarray*}
J(\lambda_0)&=&\sigma^2 \Biggl(\sum
_{g=1}^G \frac
{n_g(n_g-1)}{2}f_w(n_g)^2\\
&&\hspace*{18pt}{}+
\alpha\sum_{g=1}^G \frac
{n_g(n_g-1)}{2}
\biggl\{ \frac{n_g(n_g-1)}{2}-1 \biggr\}f_w(n_g)^2
\Biggr)
\\
&=&\sigma^2 \Biggl(\sum_{g=1}^G
1 +\alpha\sum_{g=1}^G \biggl\{
\frac
{n_g(n_g-1)}{2}-1 \biggr\} \Biggr).
\end{eqnarray*}
Thus,
\[
\frac{J(\lambda_0)}{I(\lambda_0)} =\frac{ ( G +\alpha\sum_{g=1}^G \{ {n_g(n_g-1)}/{2}-1 \}  ) }{
\sum_{g=1}^G  \{ {n_g(n_g-1)}/{2}  \}^{1/2} }.
\]

Note that this ratio does not depend on $\lambda_0$, it purely depends
on the correlation parameter $\alpha$.
We can estimate $\alpha$ from the empirical correlation of the $u(\hat
\lambda;x_i)$'s for SLV pairs within the same group. We do this formally
by modeling the score functions as multivariate Gaussian with the appropriate
covariance structure. In practice, we make the further assumption of a
common $\alpha$ value for all loci. Details are given in Appendix \ref{appb}.

\subsection{Joint inference across loci} \label{S:Joint}

It is possible to combine inferences across loci. It is natural to
assume that, conditional on parameters, the data from one locus is
independent of data from another, as the SLVs will be caused by
different recombination and
mutation events. Two natural questions to address from looking across
loci are whether the value of $\lambda$ is the same for all loci and,
if so, can we estimate this common $\lambda$ value. We will show how
composite likelihood
methods can be used to answer both these questions.

First we introduce some notation. Let $\operatorname{Cl}^{(l)}(\lambda)$ be the
composite log-likelihood function for locus $l$. Similarly, let
$J(\lambda)^{(l)}$ and $I(\lambda)^{(l)}$
denote the corresponding values of $J(\lambda)$ and $I(\lambda)$ for
locus $l$.

We first consider answering the second question, and let $\lambda_0$ be
the true common $\lambda$ value for the loci. We can estimate this by
maximizing the sum of the locus-specific composite log-likelihoods:
\[
\hat\lambda=\arg\max\sum_{l=1}^L
\operatorname{Cl}^{(l)}(\lambda).
\]
Furthermore, if we define a scaled deviance as
\[
W(\lambda)=\frac{2}{\gamma} \Biggl[\sum_{l=1}^L
\operatorname{Cl}^{(l)}(\hat \lambda) - \sum_{l=1}^L
\operatorname{Cl}^{(l)}(\lambda) \Biggr],
\]
where $\gamma=\sum_{l=1}^L J^{(l)}(\lambda_0)/\sum_{l=1}^L
I^{(l)}(\lambda_0)$, then $W(\lambda_0)$ asymptotically has a
chi-squared distribution with one degree of freedom [\citet{Varin:2011}].
We can estimate $\gamma$ using our estimates of $J(\lambda_0)^{(l)}$
and $I(\lambda_0)^{(l)}$ from each locus.

To test whether the value of $\lambda$ at each locus is the same, we
can use a (composite) likelihood-ratio test statistic. We define the
test statistic to be proportional to the difference in composite
log-likelihood for a model which
allows each locus to have different $\lambda$ values and that of a
model with a common $\lambda$ value across loci:
\[
\mathit{LR}=\frac{2}{\nu_1} \Biggl[\sum_{l=1}^L
\max_\lambda \operatorname{Cl}^{(l)}(\lambda ) - \max
_\lambda\sum_{l=1}^L
\operatorname{Cl}^{(l)}(\lambda) \Biggr].
\]
If our null hypothesis, of a common $\lambda$ across loci, is true,
then $\mathit{LR}$ has an asymptotic distribution that is an inhomogeneous sum
of independent chi-squared distributions [\citet{Varin:2011}]. For an
appropriate choice
of $\nu_1$ it is common to approximate this asymptotic distribution for
a chi-squared distribution with $L-1$ degrees of freedom [\citet
{Rotnitzky:1990,Molenberghs:2005}]. We can calculate $\nu_1$ based on
estimates of
$J(\lambda_0)^{(l)}$ and $I(\lambda_0)^{(l)}$ for
each locus. Details are given in Appendix \ref{appc}.

\section{Simulation study} \label{S:SS}

We have investigated this approach for estimating the relative rate of
recombination to mutation using simulation. We used {\texttt{simMLST}}
[\citet{Didelot:2009}] to simulate MLST data sets, under a standard
neutral model
for evolution. This involves a model for recombination where the tract
length of a recombination event is geometrically distributed. As a
default scenario we used parameter values that are appropriate for MLST
data for a range of
bacteria. This involves data at 7 loci, a population scaled mutation
rate, $\theta$, of 100 across the 7 loci, $\lambda=1$, and
mean recombination tract lengths that are 5 times the length of the
gene fragments used for each MLST locus. We then considered performance
of the method as we varied $\theta$, $\lambda$ and the number of loci.
Note that for {\texttt{simMLST}} recombination rate is defined in terms
of the rate at which any loci are affected by recombination, and hence
is equal to the product of $\lambda$ and $\theta$.


\subsection{Estimating \texorpdfstring{$\lambda$}{lambda}} \label{S:R1}

For each scenario we present results from analyzing each locus
individually (denoted \emph{Individual}) and results for a combined
analysis of data at all loci under
the assumption of a common $\lambda$ value (denoted \emph{Joint}).
Results are averaged across 100 simulated data sets for each scenario,
and further averaged across loci for the individual analysis. We
present results in terms of estimating $\lambda$,
the relative rate of recombination to mutation. In all cases we
look at the bias of the estimates, their root mean square error and the
coverage of putative 95\% confidence intervals. For each batch of
simulations we also present the average number of STs and the number of
SLVs (across all loci) per
data set. When estimating the distribution of the number of nucleotides
introduced at a recombination event (see Appendix \ref{appa}), we assumed that
with probability 0.8 a recombination event changed the complete DNA sequence
in a region.

\begin{table}
\caption{Results for estimating the relative rate of
recombination to mutation for a different number of samples, $N$, and
loci, $L$. In each case we give the mean number of STs and SLVs per
data set, the bias
and root mean square error of estimating $\lambda$, and the coverage
for putative 95\% confidence intervals for $\lambda$. All simulations
had $\lambda=1$. For the simulations with $L=7$ we fixed $\theta=100$
across the 7 loci. For simulations
with other numbers of loci we fixed $\theta=14$ per locus. \emph
{Individual} gives the results for analyzing a single-locus, \emph{joint}
for combining inferences
across all loci}
\label{Tab:Res1}
\begin{tabular*}{\textwidth}{@{\extracolsep{\fill}}ld{4.0}d{4.0}d{2.2}d{1.2}cd{2.2}cc@{}}
\hline
& & & \multicolumn{3}{c}{\textbf{Individual}} & \multicolumn{3}{c@{}}{\textbf{Joint}} \\[-6pt]
& & & \multicolumn{3}{c}{\hrulefill} & \multicolumn{3}{c@{}}{\hrulefill} \\
 & \multicolumn{1}{c}{\textbf{STs}} & \multicolumn{1}{c}{\textbf{SLVs}} & \multicolumn{1}{c}{\textbf{Bias}} &
\multicolumn{1}{c}{\textbf{RMSE}} & \multicolumn{1}{c}{\textbf{Coverage}}& \multicolumn{1}{c}{\textbf{Bias}} &
\multicolumn{1}{c}{\textbf{RMSE}} & \multicolumn{1}{c@{}}{\textbf{Coverage}}\\
\hline
${N}$ &\multicolumn{8}{c}{$L=7$} \\
\phantom{0.}5000 & 629 & 600 & -0.06 & 0.3 & 0.92 &-0.1 & 0.15 & 0.80 \\
10,000 & 758 & 797 & -0.04 & 0.26 & 0.94 &-0.06 & 0.12 & 0.85 \\
20,000 & 884 & 1000 & -0.03 & 0.25 & 0.92 &-0.06 & 0.11 & 0.87 \\
40,000 & 1013 & 1231 & -0.01 & 0.23 & 0.93 &-0.03 & 0.09 & 0.91 \\[3pt]
$L$&\multicolumn{8}{c}{$N=10\mbox{,}000$} \\
\phantom{0}5 & 574 & 730 & 0 & 0.29 & 0.92 &-0.03 & 0.12 & 0.94 \\
10 & 973 & 891 & -0.05 & 0.27 & 0.94 &-0.08 & 0.11 & 0.83 \\
20 & 1591 & 1196 & -0.07 & 0.31 & 0.92 &-0.11 & 0.14 & 0.51 \\
\hline
\end{tabular*}
\end{table}

Table~\ref{Tab:Res1} shows results as we vary the number of isolates in
our sample and the number of loci. First consider the individual
analysis. Increasing $N$ or $L$ makes only a small impact on the
quality of inference.
Note that the root mean square error does not decrease much as $L$
increases, because we are independently estimating a value of $\lambda$
for each locus and the number of SLV pairs per locus is actually
reducing. Coverage
values are close to their nominal level.

\begin{table}[b]
\caption{Results for estimating the relative rate of
recombination to mutation for different mutation and recombination
rates. We use $L=7$ and $N=10\mbox{,}000$ in all cases. In each case we give
the mean number of STs and SLVs per data
set, the bias
and root mean square error of estimating $\lambda$, and the coverage
for putative 95\% confidence intervals for $\lambda$. Individual gives
the results for analyzing a single-locus, joint for combining inferences
across all loci}
\label{Tab:Res2}
\begin{tabular*}{\textwidth}{@{\extracolsep{\fill}}ld{4.0}d{4.0}d{2.2}d{1.2}cd{2.2}cc@{}}
\hline
& & & \multicolumn{3}{c}{\textbf{Individual}} & \multicolumn{3}{c@{}}{\textbf{Joint}} \\[-6pt]
& & & \multicolumn{3}{c}{\hrulefill} & \multicolumn{3}{c@{}}{\hrulefill} \\
 & \multicolumn{1}{c}{\textbf{STs}} & \multicolumn{1}{c}{\textbf{SLVs}} & \multicolumn{1}{c}{\textbf{Bias}} &
\multicolumn{1}{c}{\textbf{RMSE}} & \multicolumn{1}{c}{\textbf{Coverage}}& \multicolumn{1}{c}{\textbf{Bias}} &
\multicolumn{1}{c}{\textbf{RMSE}} & \multicolumn{1}{c@{}}{\textbf{Coverage}}\\
\hline
${\lambda}$& \multicolumn{8}{c}{$\theta=100$} \\
0.2 & 524 & 583 & 0 & 0.08 & 0.94 &0 & 0.04 & 0.86 \\
0.5 & 612 & 665 & -0.02 & 0.15 & 0.95 &-0.03 & 0.07 & 0.90 \\
2 & 1007 & 1010 & -0.12 & 0.49 & 0.93 &-0.18 & 0.26 & 0.80 \\
5 & 1648 & 1480 & -0.39 & 1.4 & 0.9 &-0.59 & 0.73 & 0.72 \\[3pt]
${\theta}$& \multicolumn{8}{c}{$\lambda=1$} \\
\phantom{0}20 & 188 & 236 & 1.92 & 9.22 & 0.93 &-0.05 & 0.29 & 0.93 \\
\phantom{0}50 & 426 & 485 & -0.03 & 0.45 & 0.92 &-0.09 & 0.17 & 0.84 \\
200 & 1273 & 1218 & -0.05 & 0.19 & 0.93 &-0.06 & 0.1 & 0.82 \\
500 & 2374 & 1908 & -0.03 & 0.14 & 0.94 &-0.04 & 0.07 & 0.87 \\
\hline
\end{tabular*}
\end{table}

Combining information across loci gives more accurate estimates in all
cases as measured by root mean square error; however, coverage
proportions drop substantially below the nominal level in many cases.
We believe this is
because of a slight bias in our estimates, due to the approximate
model that we are fitting. The impact of this bias is seen more
strongly when the uncertainty of the estimates is lower, such as when
combining information
across multiple loci. For our simulations we can test whether this is
the case, because we are able to ``cheat'' and remove the bias and then
see if the resulting confidence intervals have appropriate coverage
probabilities.
We remove the bias by multiplying our estimate
of $\lambda$ by an appropriate constant, chosen so that this new
estimator is unbiased. If we do this, coverage values for the joint
analysis for all scenarios we considered were in the range 93\% to 97\%, which is consistent with the
nominal significance level, suggesting that we are assessing correctly
the uncertainty in our estimators.

We also looked at how the parameters for the mutation and recombination
rate affected performance. These are given in Table~\ref{Tab:Res2}. We
see that for a fixed mutation rate, we get similar performance for
a range of $\lambda$ values. Note that while the bias and root mean
square error is increasing as $\lambda$ increases, this is because we
are estimating a larger value: the relative size of bias and error
remains roughly constant. As we vary $\theta$ we notice that
performance gets better as $\theta$ increases. The large root mean
square error values for small $\theta$ are caused by estimating a
ratio, and the distribution of
the estimator in these cases is highly skewed with occasional large
estimates for $\lambda$. Again, we get more accurate estimates when we
combine information across loci, but the coverage values drop
noticeably below their
nominal level. As mentioned above, this is due to slight bias in the
model, which has greater impact for the joint analysis.

\subsection{Testing for variation in \texorpdfstring{$\lambda$}{lambda}}

We further looked at both type I error rate and power for testing for a
difference in $\lambda$ across loci. For all the scenarios presented in
Section~\ref{S:R1} we implemented our test for detecting a difference
in $\lambda$ values
across loci. We implemented the test at the 95\% significance level.
These scenarios all correspond to the case where there is a common
$\lambda$ value. The average type I error rate across these scenarios
was 7\%, with a
range of 3\% to 12\%. There was no obvious pattern to which scenarios
had higher, or lower, type I error rates, and the range of values
observed across the scenarios is consistent with the random
fluctuations one would
expect if the type I error rate was the same in all cases.


\begin{table}
\tablewidth=200pt
\caption{Power of test for detecting variation in
$\lambda$ across loci. In all cases we simulated data of sample size
10,000 at 7 loci, with $\rho=20$ per locus. In scenario \textup{(a)} we have one
locus with $\theta=20$
and all others with $\theta=20/C$; in scenario \textup{(b)} we have one locus
with $\theta=20/C$ and all other scenarios with $\theta=20$}
\label{Tab:Power}
\begin{tabular*}{200pt}{@{\extracolsep{\fill}}lcc@{}}
\hline
$\bolds{C}$ & \textbf{(a)} & \textbf{(b)} \\
\hline
2 & 0.46 & 0.33 \\
3 & 0.93 & 0.64 \\
4 & 0.97 & 0.71 \\
\hline
\end{tabular*}
\end{table}

We then investigated the power of the test. We simulated data at 7
loci, with the population-scaled rate at which each locus is affected
by recombination being $20$, but with different mutation rates per locus.
We tried two sets of scenarios: (a)~where one locus had $\theta=20$
and the
others had $\theta=20/C$; and (b)~where one locus had $\theta=20/C$ and
all others had $\theta=20$. The first scenario is where most loci have
$\lambda=C$, but one locus has $\lambda=1$; for the second
scenario this is reversed. We repeated this for $C=2,3,4$, reflecting
different levels of variation in $\lambda$ across the loci. Such data
sets can be simulated using {\texttt{simMLST}} with
$\theta=20$ per locus and $\lambda=1$, and then
using thinning (removing mutations at a proportion of sites) to reduce
the mutation rate at some loci. It is not possible to use {\texttt
{simMLST}} to generate data with the same $\theta$ but different $\rho$
per locus.

Results are given
in Table~\ref{Tab:Power}.
The results suggest large power for detecting variation in $\lambda$ of
a factor of three or more in scenario (a). There is less power under
scenario (b), as there are fewer mutations at the one locus for which
$\lambda$ is
different to the others, and this makes it harder to detect the
difference in $\lambda$.

\subsection{Comparison with a bootstrap approach}

Finally, we give a simple comparison with a parametric bootstrap
approach to calculating confidence intervals. The aim here is to give
some insight into the challenges and issues relating to the use of the
bootstrap,
over and above the extra computational cost it incurs. To mimic issues
that occur in real data, we will use a different model to simulate the
data than that which we assume when implementing the bootstrap. Our simulated
data was generated by {\texttt{simMLST}} under a model where there had
been recent population growth. If we define the population size, $N$,
say, to be the size of population prior to the growth, then our model
assumes a step
change in the population size from $N$ to $10N$ at a time $0.1$ in the
past. The population-scaled mutation rate is 100 across the 7 loci, and
$\lambda=1$. The effect of this model is to reduce the relative rate
of coalescent to mutation and recombination for the larger population
size, and hence increase the overall number of mutation and
recombination events near the tips of the clonal frame.
We simulated 100 data sets, each with sample size of 1000 isolates. A
summary of the results from analyzing these data sets using the
composite likelihood method is shown in Table~\ref{Tab:Rob}. The
observed type I error
rate for the test of a common $\lambda$ across loci at $95\%$
significance level was 0.03.

\begin{table}[b]
\caption{Results for estimating relative rate of
recombination to mutation for a population growth scenario. We give
coverage for three types of putative 95\% confidence intervals: those
from asymptotic theory
for composite likelihood, and two parametric bootstrap approaches. The
latter differ in whether they simulate data sets that match the true
data in terms of the number of STs or the number of SLVs. Results are
shown for both
analyzing loci individually and a joint analysis}
\label{Tab:Rob}
\begin{tabular*}{\textwidth}{@{\extracolsep{\fill}}ld{2.2}cccc@{}}
\hline
& \multicolumn{5}{c@{}}{\textbf{Population growth (Average 810 STs 173 SLVs)}} \\[-6pt]
& \multicolumn{5}{c@{}}{\hrulefill} \\
& \multicolumn{3}{c}{\textbf{Composite likelihood}} & \multicolumn{2}{c@{}}{\textbf{Bootstrap}} \\ [-6pt]
& \multicolumn{3}{c}{\hrulefill} & \multicolumn{2}{c@{}}{\hrulefill} \\
& \multicolumn{1}{c}{\textbf{Bias}} & \textbf{RMSE} & \textbf{Coverage}& \textbf{Coverage (ST)} & \multicolumn{1}{c@{}}{\textbf{Coverage (SLV)}}\\
\hline
Individual & 0.03 & 0.56 & 0.95 & 0.69 & 0.98\\
Joint & -0.08 & 0.17 & 0.93 & 0.65 & 0.99\\
\hline
\end{tabular*}
\end{table}

To implement a parametric bootstrap, we then simulated data from a
constant population size model. We fixed the population-scaled mutation
rate to be equal to the average number of the difference between a pair
of isolates, and set
$\lambda=1$. Data sets simulated from a constant population size model
have different patterns in terms of the ratio of SLVs to STs that are
observed for the population growth model.
Thus, we considered two approaches to simulating data for the
parametric bootstrap. For each ``real'' data set we analyzed we
simulated 100 data sets under the constant population size model. Our
first approach simulated
each of these 100 data sets to have the same number of STs as the real
data set, while the second approach matched in terms of the number of
SLVs. The use of the parametric bootstrap added considerably to
the cost of analyzing the data. The composite likelihood approach takes
a matter of seconds to run, as compared to of the order of 10 hours to
simulate 100 SLV data sets.

To construct our parametric bootstrap confidence interval, we used the
approach of \citet{Yu:2012}. This is based on noting that the sampling
distribution of an estimate of $\lambda/(1+\lambda)$ is approximately normal.
We used the 100 data sets simulated under the parametric bootstrap to
estimate the variance of this normal distribution. This enables us to
produce putative 95\% symmetric confidence intervals for $\lambda
/(1+\lambda)$, which
can then be transformed to confidence intervals for $\lambda$.

The observed coverage for each of these two methods for constructing
bootstrap confidence intervals is shown in Table~\ref{Tab:Rob}. We
observe that choosing the size of the data sets simulated by the
parametric bootstrap through
matching the number of STs leads to much smaller confidence intervals
than when we match on the number of SLVs. The coverage for the
intervals when we match on STs is substantially lower than the putative
95\% confidence
level. In this example matching on SLVs gives much better results,
though the coverage results suggest that the confidence intervals
produced are slightly too conservative.

The main message from these simulations is that when performing a
parametric bootstrap there can be issues if features of the simulated
data sets do not match the real data set when these features affect the
amount of information the data has about the parameter of interest.
These issues disappear if we are able to simulate from a model which is
a very good approximation to real life, but for bacterial data this
is rarely the case. 

\section{Application to bacterial MLST data} \label{S:Bac}

We applied our composite likelihood method to detect variation in
$\lambda$ across loci
and estimate $\lambda$ for a range of bacteria. We used MLST data
downloaded from {\url{http://pubmlst.org/}}. In each case
we had data at 7 loci. The
bacteria we considered, together with the number of SLV pairs we
obtained, were as follows:
\emph{Bacillus cereus} (281 SLV pairs); \emph{Enterococcus faecium}
(481); \emph{Haemophilus influenzae} (977);
\emph{Klebsiella pneumoniae} (404); \emph{Staphylococcus aureus} (7892);
\emph{Streptococcus uberis} (356);
\emph{Campylobacter jejuni} (7417); and \emph{Campylobacter coli} (1842).

\begin{figure}

\includegraphics{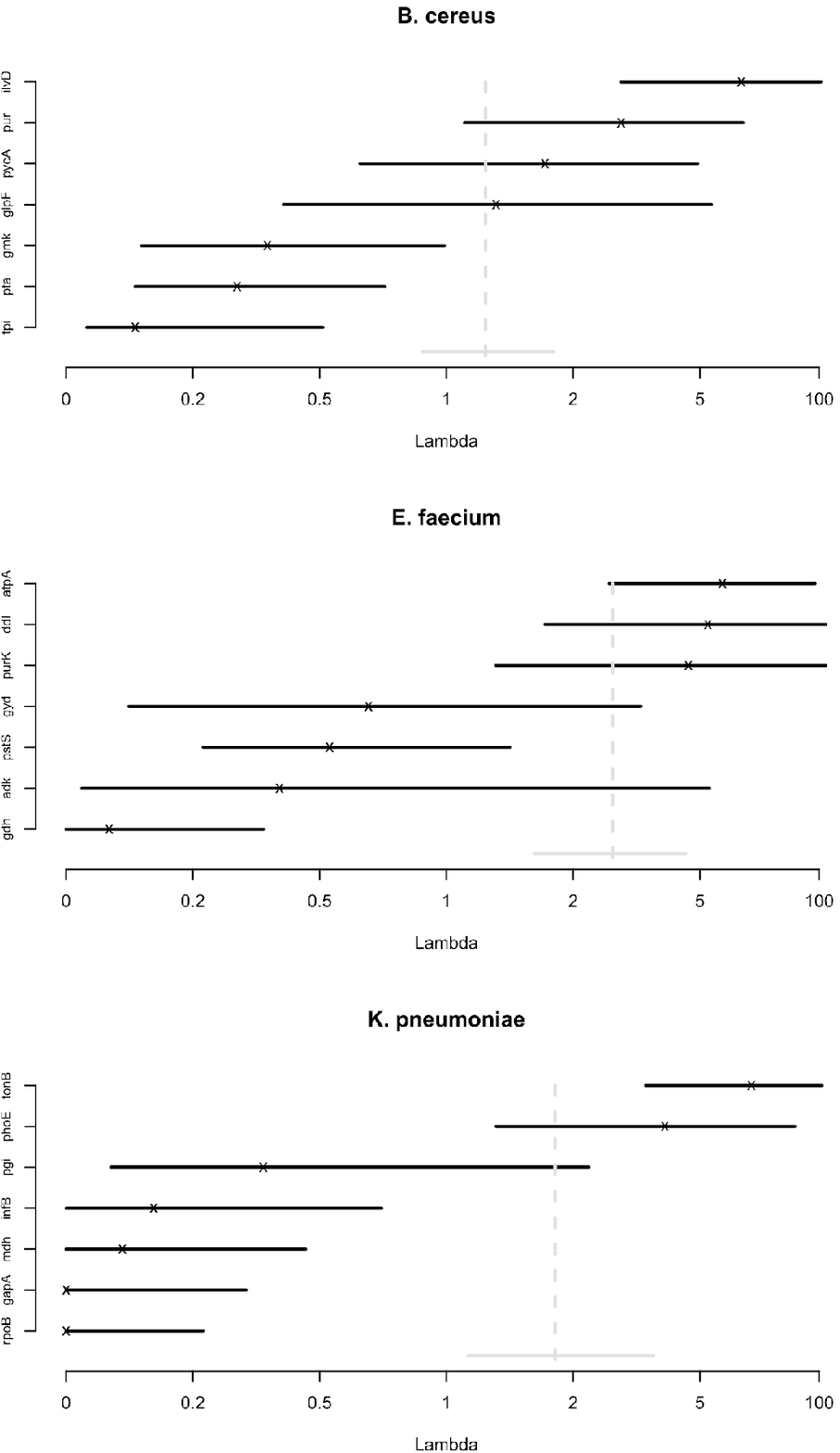}

\caption{Estimates and confidence intervals for
$\lambda
$ for the three bacteria that showed evidence of variation across loci.
For each bacteria we plot the estimate (cross) and putatitve 95\% confidence
intervals (lines) for each locus. In gray is the estimate (vertical
dashed line) and 95\% confidence interval (horizontal line) under an
assumption of a common value of $\lambda$ across loci. We have ordered
the loci
in terms of the value of the estimate of $\lambda$, with decreasing
estimates as we move down each plot. For clarity we have chosen an
$x/(1+x)$ scale for the $x$-axis for each plot.}\label{Fig:B1}
\end{figure}

We found evidence for variation of $\lambda$ across loci ($p$-values less
than 0.01) in 3~bacteria: \emph{B. cereus} ($p$-value $2.2\times10^{-8}$);
\emph{E. faecium} ($9.4\times10^{-6}$); and \emph{K. pneumoniae}
($4.9\times10^{-12}$). Estimates of $\lambda$ for each locus for these
bacteria are given in Figure~\ref{Fig:B1}, and estimates of a common
$\lambda$ for
the other bacteria are given in Table~\ref{Tab:B2}.

\begin{table}
\tablewidth=200pt
\caption{Estimate of common $\lambda$ across MLST loci,
together with putative 95\% confidence
intervals}
\label{Tab:B2}
\begin{tabular*}{200pt}{@{\extracolsep{\fill}}ld{1.2}c@{}}
\hline
\textbf{Bacteria} & \multicolumn{1}{c}{\textbf{Estimate of} $\bolds{{\lambda}}$} & \multicolumn{1}{c@{}}{\textbf{95\% CI}} \\
\hline
\emph{H. influenzae} & 4.9 & $(3.3,7.4)$ \\
\emph{S. aureus} & 1.4 & $(0.92,2.1)$\\
\emph{S. uberis} &11 & $(4.8,180)$\\
\emph{C. jejuni} &3.4 &$(2.9,4.1)$ \\
\emph{C. coli} & 0.43 &$(0.21,0.88)$ \\
\hline
\end{tabular*}
\end{table}

The most striking results are the degree of variation we see in
estimates of $\lambda$ for \emph{B. cereus}, \emph{E. Faecium} and
\emph{K. pneumoniae}. For each bacteria we have evidence for loci with
$\lambda<1$, and often
$\lambda\approx0$, as well as for loci with $\lambda$ substantially
greater than 1. The estimates we get are consistent with the relative
rate of recombination to mutation varying by between one and two orders of
magnitude across the loci for each bacteria.

For the five bacteria for which we see no evidence for variation in
$\lambda$, we see that the relative rate of recombination to mutation
varies noticeably across the bacteria. The order of bacteria from
the one with highest to lowest $\lambda$ is \emph{S. uberis}, \emph{H.
influenzae}, \emph{C. jejuni}, \emph{S. aureus} and \emph{C. coli}. The
estimates of $\lambda$ for \emph{S. uberis} and \emph{C. coli}
differ by
a factor of 25.

\section{Conclusion}

We have presented a way of both estimating the relative rate of
recombination to mutation and detecting recombination rate variation from
MLST data. The key novelty within the statistical approach we take is
to directly model the form of dependence of information from
different SLVs. This enables us to correct for the dependence between
the contribution each SLV makes to the composite likelihood, and thus get
appropriate measures of uncertainty in the estimates of parameters we
get from maximizing the composite likelihood, and also
to test for variation in recombination rate across loci. While
composite likelihood methods
are extensively used within genetics [see, e.g., \citet
{Larribe:2011} and references therein], to date they have been
primarily used to get point estimates of parameters. Our results
suggest that, with appropriate modeling of the dependence structure, it
should be possible to extend these earlier methods to obtain both point
estimates
and confidence intervals for parameters of interest.

An example of why this is important can be seen by a previous analysis
of MLST data in bacteria. Recombination rates in \emph{B. cereus} at
MLST loci were estimated in \citet{PerezLosada:2006},
using an alternative composite likelihood approach [\citet
{Hudson:2001,McVean/Awadalla/Fearnhead:2002}]. They
observed estimates of the recombination rate varying by a factor of
nearly 50 across the loci; however, they were unable to calculate
confidence intervals for their recombination rate estimates.
As a result, it was impossible to conclude whether this variation is
due to variability in the estimator or whether it resulted from
true differences in recombination rates across the loci. By comparison,
our analysis gives strong evidence for variation in $\lambda$ across
these loci.

Recently MLST data are increasingly being replaced by full-genome data.
The methods developed here can still be applied to such data, by
choosing a set of $L$ loci and summarizing
the full-genome data in terms of SLV pairs and the nucleotide
differences of each pair. To be viable, such an approach would need
full-genome data from a substantial number
of isolates in order to produce sufficient SLV pairs. Summarizing the
data in such a way would clearly lead to a substantial loss of
information, but would be a simple and quick
approach to performing an initial analysis of data as compared to
methods that try and analyze the full sequence data [e.g., \citet
{Didelot:2010}]. As pointed out by a reviewer,
the flexibility over the choice of loci would give the possibility of
using a nonparametric bootstrap to assess uncertainty in estimates. If
interest is in the ratio
of recombination to mutation at a given loci, we can make different
choices for the other $L-1$ loci. Each choice would give a different
set of SLVs, and hence a different estimate.
The variability of these estimates could be used to measure the degree
of uncertainty in the final estimate we make.

The results from our application to data from eight bacteria species
are in line with results in \citet{Vos/Didelot:2009}. While that
paper estimated a different measure of recombination to mutation,
looking at the probability of a nucleotide change as due to
recombination rather
than mutation, the ordering of bacteria species from less to more
recombinant is broadly similar.
The more striking results, though, relate to strong evidence of rate
variation in the rate of recombination to mutation in three of the
bacteria. This is part of
the growing evidence for substantial recombination rate variation, for
example, in \citet{Didelot:2010}, who also found evidence of
rate variation in \emph{B. cereus}, and \citet{Guy:2012}, who observed 3
orders of magnitude of recombination rate variation in
\emph{Bartonella henselae}. More indirect evidence for rate variation
also comes from the variation in recombination rates for closely related
bacterial species and substantial differences in estimates of
recombination rates from different studies of a given bacterial species
[see \citet{Didelot/Maiden:2010} for more details].

The reasons behind substantial variation in the relative rate of
recombination to mutation are currently unclear. One explanation is
that estimated recombination rates are higher
within regions under positive selection [\citet{Vos:2009}]. The argument
is that we are only likely to see recombination events that add beneficial
or remove deleterious mutations. The selective advantage of such
recombination events over mutation will be largest within genes for which
selection is strongest. In our study we see large variation in
recombination rates among housekeeping genes, genes we would expect to
all be under
strong selective pressure. This includes variation in recombination
rates between genes with similar function: for example, both \emph{pycA}
and \emph{tpi} in
\emph{B. cereus} are genes involved in gluconeogenesis, yet their
estimates of $\lambda$ differ by an order of magnitude. This suggests
that there are other
factors responsible for the variation that we observe.



The MLST data analyzed in Section~\ref{S:Bac}, together with {\texttt
{R}} code implementing the composite likelihood method presented in
this paper,
are available from \url{http://www.maths.lancs.ac.uk/\textasciitilde fearnhea/SLV.zip}.
\begin{appendix}
\section{Estimating \texorpdfstring{$\Pr_\lambda(X=x|\mbox{SLV},A^c)$}{Prlambda(X=x|SLV,Ac)}}\label{appa}

Our approach to approximating $\Pr_\lambda(X=x|\mbox{SLV},A^c)$ for a
given locus is to use a Monte Carlo estimate of the probability of $x$
nucleotide differences being imported
at a single recombination event. This simple idea is based upon the
fact that for an SLV pair we expect the isolates to have a recent
common ancestor, and hence the
number of mutation/recombination events to be one with a high
probability. It also simplifies computations in that this approximation
is independent of $\lambda$,
and hence can be calculated and stored once. It is possible to extend
the following Monte Carlo setup to allow for possible multiple events
at the locus, but to do
this correctly would involve making the approximation of $\Pr_\lambda
(X=x|\mbox{SLV},A^c)$ depend on $\lambda$.

We assume we have a sample of $K$ isolates, and for each pair $(i,j)$
know the number of nucleotide differences at locus $l$ between isolates
$i$ and $j$, denoted $x_{ij}$.
Let $m$ be the number of bases at the SLV locus.
Assume 
with probability
$p_a$ the region of the recombination event will include all $m$ bases
of the locus. Fix the Monte Carlo sample size, $M$:
\begin{enumerate}[(1)]
\item[(1)] Set $n_i=0$ for $i=0,\ldots,m$.
\item[(2)] Repeat $M$ times:
\begin{enumerate}[(a)]
\item[(a)] Sample $i$ and $j$ independently from $\{1,2,\ldots,K\}$.
\item[(b)] With probability $p_a$ set $x=x_{ij}$; otherwise sample $u$,
a realization of a standard uniform random variable, and $x$ the
realization of a Binomial random
variable with parameters $x_{ij}$ and $u$.
\item[(c)] Set $n_x=n_x+1$.
\end{enumerate}
\item[(3)] Calculate the approximation
\[
\Pr_\lambda\bigl(X=x|\mbox{SLV},A^c\bigr)\approx
\frac{n_x+1}{M+m-n_0}.
\]
\end{enumerate}
In step (2b) we have used a simple mechanism for simulating the number
of changes due to a recombination event for which a breakpoint lies
within the locus. This involves simulating $u$, the proportion
of the locus that is affected by the recombination event, and then,
conditional on this, how many nucleotide differences the recombination
event introduces.

The final approximation used in part (3) is chosen so that all
probabilities are non 0, to allow for the possibility of a value $x$
for the number of nucleotide differences observed for an SLV pair
that was not simulated. We subtract $n_0$ from the denominator, as we
are conditioning on there being an SLV pair at locus $l$, in which case
$x\neq0$. We repeat this procedure to get
a different distribution for the number of nucleotide differences for
each locus.

\section{Estimating \texorpdfstring{$\alpha$}{alpha}}\label{appb}

Assume we have partitioned the STs in $G$ groups of size $n_1,\ldots
,n_G$, and for each SLV pair we have the value of the score function at
$\hat\lambda$. To estimate the within-group correlation of the score
statistics, we will model the scores as being Gaussian, with
independence across groups. Within
group $g$ we will view the scores as realization of a vector
random-variable $V=(u(\lambda_0;X_1),\ldots,u(\lambda_0;X_k))$, where
$k=n_g(n_g-1)/2$. We model
$V\sim\operatorname{MVN} (\mathbf{0},\Sigma )$, and $\Sigma$ is a
$k\times k$ variance--covariance matrix for $i=1,\ldots,k$, with
$\Sigma
_{ii}=\sigma^2$; and for $i,j=1,\ldots,k$ with $i\neq j$, $\Sigma
_{ij}=\alpha\sigma^2$. If we denote the data for this group as
$v=(u(\hat\lambda;x_1),\ldots,u(\hat\lambda;x_k))$, then the likelihood
for the group is
$ l(\alpha,\sigma;v)= -0.5\log\operatorname{det}(\Sigma)-\frac
{1}{2}v\Sigma
^{-1} v^T$.
Using Sylvester's theorem,
\[
\operatorname{det}(\Sigma)=\sigma^{2k}(1-\alpha)^k \biggl(
\frac
{1+(k-1)\alpha
}{1-\alpha} \biggr) \quad\mbox{and}\quad \Sigma^{-1}=a_kI_k+b_k1_k,
\]
where $I_k$ is the identity matrix, $1_k$ is a $k\times k$ matrix of
1's, $a_k=1/(1-\alpha)$ and
$ b_k={-\alpha}/{(1-\alpha)[1+(n-1)\alpha]}$.
Thus,
\begin{eqnarray*}
l(\alpha,\sigma;v)&=& -\frac{k}{2}\log\bigl(\sigma^2[1-\alpha]
\bigr)-\frac
{1}{n}\log \biggl(\frac{1+(k-1)\alpha}{1-\alpha} \biggr)\\
&&{} -
\frac{1}{2} \Biggl(a_k\sum_{i=1}^k
u(\hat\lambda;x_i)^2+b_k \Biggl[ \sum
_{i=1}^k u(\hat\lambda;x_i)
\Biggr]^2 \Biggr).
\end{eqnarray*}
If we denote the data in group $g$ by $x^{(g)}_1,\ldots,x_{k_g}^{(g)}$
where $k_g=n_g(n_g-1)/2$, then we get a likelihood
\begin{eqnarray*}
&&\sum_{g=1}^G \Biggl\{ -
\frac{k_g}{2}\log\bigl(\sigma^2[1-\alpha]\bigr)-
\frac
{1}{n}\log\biggl(\frac{1+(k_g-1)\alpha}{1-\alpha} \biggr) \\
&&\qquad{}- \frac{1}{2}
\Biggl(a_{k_g}\sum_{i=1}^{k_g} u
\bigl(\hat\lambda ;x^{(g)}_i\bigr)^2+b_{k_g}
\Biggl[ \sum_{i=1}^k u\bigl(\hat\lambda
;x^{(g)}_i\bigr) \Biggr]^2 \Biggr)\Biggr\},
\end{eqnarray*}
which we maximize numerically over $\sigma>0$ and $\alpha\in[0,1)$ to
get an estimate of~$\alpha$. In practice, we use a common value of
$\alpha$ for all loci, obtained by averaging the locus-specific
estimates.

\section{Estimating \texorpdfstring{$\nu_1$}{nu1}}\label{appc}

Consider a model for data at $L$ loci. The general model will have
parameter vector $(\lambda_1,\ldots,\lambda_L)$ for the value of the
rate of recombination to mutation for each of the $L$ loci. Under
our assumption of independence across loci, our joint composite
log-likelihood is
$ \operatorname{Cl}^*(\lambda_1,\ldots,\lambda_L)=\sum_{l=1}^L \operatorname{Cl}^{(l)}(\lambda_l)$,
the sum of the composite log-likelihoods for each locus.

Assume that there is a common $\lambda$ value for all loci. Let
$\lambda
_0$ denote the true common value. Further, to simplify notation, let
$J^{(l)}=J^{(l)}(\lambda_0)$ and $I^{(l)}=I^{(l)}(\lambda_0)$ be the value
of $J$ and $I$ at locus $l$ evaluated at this true common value. Then
the $J$ and $I$ matrices associated with our joint composite
log-likelihood will be diagonal with entries $(J^{(1)},\ldots
,J^{(L)})$ and
$(I^{(1)},\ldots,I^{(L)})$, respectively.

We are interested in a test for whether there is a common $\lambda$
value for all loci. To do this, we can introduce a reparameterization
to $(\phi_1,\ldots,\phi_L)$, where $\phi_1=\lambda_1$, and for
$l=2,\ldots,L$, $\phi_l=\lambda_l-\lambda_1$.
So a common $\lambda$ value is equivalent to $\phi_l=0$ for
$l=2,\ldots
,L$. Let
\[
\operatorname{Cl}_\phi(\phi_1,\ldots,\phi_L)=\operatorname{Cl}^*(
\phi_1,\phi_2+\phi_1,\ldots ,\phi
_L+\phi_1)
\]
be the composite log-likelihood under this parameterization. The
corresponding $J$ and $I$ matrices will be denoted by $J_\phi$ and
$I_\phi$.
The diagonal, first row and column of $J_\phi$ are $(\sum_{l=1}^L
J^{(l)},J^{(2)},J^{(3)},\ldots,J^{(L)})$, with all other entries being 0;
and $I_\phi$ depends on $I^{(1)},\ldots,I^{(L)}$ in a similar way.

The likelihood ratio statistic for testing $\phi_l=0$ for $l=2,\ldots
,L$ is given by
\begin{eqnarray*}
\mathit{LR}^*&=& 2 \bigl[\max \operatorname{Cl}_\phi(\phi_1,\ldots,
\phi_L) - \max \operatorname{Cl}_\phi (\phi _1,0,\ldots,0)
\bigr]\\
& =& 2 \Biggl[\sum_{l=1}^L\max
\operatorname{Cl}^{(l)}(\lambda) - \max\sum_{l=1}^L
\operatorname{Cl}^{(l)}(\lambda) \Biggr].
\end{eqnarray*}
Define $H$ and $G$ to be $(L-1)\times(L-1)$ matrices obtained from
removing the first row and column from $I_\phi^{-1}$ and $ (I_\phi
J_\phi^{-1} I_\phi)^{-1} $, respectively. Let $\eta_i$ be the
eigenvalues of the
matrix $H^{-1}G$. Then
if $\phi_l=0$ for $l=2,\ldots,L$, the asymptotic distribution of $\mathit{LR}^*$
is $\sum_{i=1}^{L-1} \eta_i Z_i^2$, where $Z_1,\ldots,Z_{L-1}$ are
independent standard normal random variables [see \citet{Varin:2011,Kent:1982}].

We approximate this distribution by scaling $\mathit{LR}^*$ to have the same
mean as a chi-squared distribution with $L-1$ degree of freedom, $\chi
^2_{L-1}$ [\citet{Rotnitzky:1990}].
Thus, we define $\nu_1=\sum_{i=1}^{L-1} \eta_i/(L-1)$, set
$\mathit{LR}=(1/\nu
_1) \mathit{LR}^*$, and approximate the distribution of $\mathit{LR}$ by a $\chi^2_{L-1}$
distribution. Higher order approximations are possible [\citet
{Varin:2008}], but we did not find them
to be more accurate in practice.
\end{appendix}

\section*{Acknowledgments}
This publication made use of the PubMLST website (\url{http://pubmlst.org/})
developed by Keith Jolley
[Jolley and Maiden (\citeyear{JoMa10})] and sited at the
University of Oxford. The development of this site has been funded by
the Wellcome Trust.


%

%



\printaddresses
\end{document}